\documentclass{article}
\usepackage{frascatiphys}
\usepackage{graphicx}
\usepackage{amsmath}
\usepackage{amsfonts}
\usepackage{multirow}

\begin{document}
\title{ 
A COSMOGRAPHIC OUTLOOK ON DARK ENERGY AND MODIFIED GRAVITY
}
\author{
Salvatore Capozziello        \\
{\em Dipartimento di Fisica ``E. Pancini", Universit\`a di Napoli ``Federico II",} \\
{\em Via Cinthia 9, 80126 Napoli, Italy.} \\
Rocco D'Agostino     \\
{\em Scuola Superiore Meridionale, Largo S. Marcellino 10, 80138 Napoli, Italy. }
}
\maketitle
\baselineskip=11.6pt

\begin{abstract}

The cosmographic technique is a powerful model-independent tool for distinguishing between competing cosmological scenarios.
The key strengths and weaknesses of standard cosmography are discussed in view of healing the convergence problem endangering the high-redshift expansions of cosmological distances.
We focus especially on rational cosmographic approximations to reconstruct the dark energy behaviour under the $f(R)$, $f(T)$ and $f(Q)$ gravity frameworks.
Based on observational constraints over the cosmographic series, we investigate the origin of cosmic acceleration and the possibility of going beyond the standard cosmological model to explain the dark energy problem. 

\end{abstract}
\baselineskip=14pt

\section{Introduction}

Our understanding of the cosmos has significantly changed since the discovery of accelerated expansion shown by the light coming from most distant Supernovae\cite{Riess98,Perlmutter99}.
Indeed, observations showed the presence of a cosmological constant with negative pressure that accelerates the cosmic expansion at recent times\cite{Peebles03}.
The standard $\Lambda$CDM model is the most widely accepted theory that incorporates the cosmological constant effects and includes cold dark matter and baryons within a a spatially flat geometry\cite{Planck18}.
In spite of its effectiveness, the $\Lambda$CDM paradigm is characterized by a number of flaws that are mostly related to the cosmological constant problem, arisen from the challenges of reconciling the standard model of particle physics with cosmological data\cite{Weinberg89,Carroll01}. 
Because of the enigmatic feature of dark energy, some authors have looked for a different explanation for the cosmic speed-up.
Possible alternatives include taking into account that  the accelerated expansion may be due to dynamical scalar fields\cite{Copeland06}, or to the presence of a single cosmic fluid endowed with an equation of state that causes it to behave like dark matter and dark energy at high and low densities, respectively\cite{Anton-Schmidt}.

Furthermore, it has been explored the possibility to solve the dark energy problem without introducing exotic components into the energy-momentum tensor. Such scenarios refer to extensions or modifications of Einstein's gravity aimed at solving the $\Lambda$ problems from first principles\cite{Capozziello11,Nojiri11}.
In fact, corrections to the Einstein-Hilbert gravitational action have attracted a lot of attention due to their ability to provide, within a single picture, an alternative interpretation of vacuum energy and an explanation for dark matter by means of geometrical effects.
The recent evidence supporting the Starobinsky model of inflation\cite{Planck_inflation18} has renewed interest for $f(R)$ models\cite{Starobinsky07,Sotiriou10}, whose action is described by a general function of the Ricci scalar, $R$.
A different way to characterize the gravitational interaction is to consider spacetime twisted by torsion. The teleparallel description of gravity has lately attracted significant interest among all the ideas explored to explain the late-time cosmic expansion\cite{Bengochea09}. 
Hence, theories in which the torsion scalar is replaced by a nonlinear function $f(T)$\cite{Linder10,Cai16}  provide a workable framework inspired by $f(R)$ gravity. 
Even more recently, the possibility to consider non-metricity as the mediator for the gravitational interaction, while assuming vanishing curvature and torsion, has induced to investigate $f(Q)$ theories\cite{Jimenez18,Jarv18} to obtain new insights into the universe's acceleration resulting from the implication of a different geometric setup with respect to the more common Riemann geometry.

Distinguishing among different models invoked to explain the late evolution of the universe becomes therefore crucial. In this respect, model-independent approaches, such as cosmography, represent a powerful tool to discriminate between dark energy and modified gravity scenarios\cite{Rocco_review,proceeding_Polonia}.
However, the difficulty of treating high-redshift data due to the scarcity of accurate measurements at $z>1$ puzzles the use of the standard cosmographic method relying on Taylor series. Indeed, the convergence issues inherent to the short convergence radius of the Taylor series may limit the predictability of cosmography, requiring then to explore new techniques that could allow to extend the standard kinematic procedure to high redshifts. This is the case of rational approximations, based on Pad\'e and  Chebyshev polynomials\cite{chebyshev}.

In this paper, we provide an updated outlook on the most recent developments regarding the cosmographic method and its application under different theoretical frameworks to reconstruct the gravitational action and, thus, deduce the nature of the dark energy behaviour.

\section{Modern cosmography}

Cosmography is probably the most basic of all model-independent techniques.
It uses the cosmological principle's observational premise and is based on Taylor expansions of observables that could be directly compared to data. 
In principle, cosmography is a strong tool for breaking the degeneracy among cosmological models. In the context of Friedmann-Lema\^itre-Roberston-Walker (FLRW) spacetime, 
the idea is to expand the cosmic scale factor, $a(t)$, in the Taylor series around the current time, $t_0$.
This method allows to study the kinematics of the universe by means of the $a(t)$ derivatives, which provides the so-called cosmographic series\cite{Cattoen07}:
\begin{equation}
H(t)\equiv \frac{1}{a}\frac{da}{dt} , \quad q(t)\equiv -\frac{1}{aH^2}\frac{d^2a}{dt^2}, \quad j(t) \equiv \frac{1}{aH^3}\frac{d^3a}{dt^3}, \quad s(t)\equiv\frac{1}{aH^4}\frac{d^4a}{dt^4}\ ,   
\end{equation}
known as the Hubble, deceleration, jerk and snap parameters, respectively.
One can then use the above definitions to expand the luminosity distance in terms of the current values of the cosmographic parameters. In particular, for a spatially flat universe, we find
\begin{align}
d_L(z)=&\ H_0^{-1}\Big[z +  \frac{1}{2}(1 - q_0) z^2 - \frac{1}{6}(1 - q_0 - 3 q_0^2 + j_0) z^3 + \frac{1}{24}(2 - 2 q_0 \nonumber \\
& \hspace{1cm} -15 q_0^2 - 15 q_0^3 + 5 j_0 +10 q_0 j_0 s_0) z^4 +\mathcal{O}(z^5)\Big]\ ,
\label{eq:dL}
\end{align}
leading to the Hubble expansion series
\begin{equation}
H(z)\simeq H_0 \left[1 + z (1 + q_0) + \frac{z^2}{2} (j_0 - q_0^2) - \frac{z^3}{6} \left(-3 q_0^2 - 3 q_0^3 + j_0 (3 + 4 q_0) + s_0\right)\right].
\label{eq:H_Taylor}
\end{equation}

Despite the simplicity and immediate applicability of the cosmographic technique, unfortunately, the absence of numerous and very accurate data at high redshifts prevents from univocally bound the higher-order terms of the cosmographic series, weakening severely the ability to disentangle modified gravity theories from effective dark energy models. In short, the standard formulation of cosmography is affected by two major problems: first, the presence of systematic errors caused by the chosen truncation order; second, the reduced predictive power when analyzing data beyond $z=1$, exceeding the radius of convergence of the Taylor series\cite{Dunsby16}.

\subsection{Pad\'e polynomials}

In order to overcome the aforementioned issues, a intriguing possibility is to take into account rational polynomials. One first relevant example consists in using Pad\'e polynomials to approximate cosmological distance measures\cite{Aviles14}. The $(n,m)$  Pad\'e approximation of a generic function $f(z)$ is defined as
\begin{equation}
P_{n,m}(z)=\frac{\displaystyle{\sum_{i=0}^{n}a_i z^i}}{1+\displaystyle{\sum_{j=1}^{m}b_j z^j}}\,,
\label{def Pade}
\end{equation}
where the coefficients $a_i$ and $b_i$ can be found from the following system:
\begin{equation}
\left\{
\begin{aligned}
&a_i=\sum_{k=0}^i b_{i-k}\ c_{k} \ ,  \\
&\sum_{j=1}^m b_j\ c_{n+k+j}=-b_0\ c_{n+k}\ , \hspace{0.5cm} k=1,\hdots, m \ .
\end{aligned}
\right .
\end{equation}
where $c_k$ are the coefficients of the Taylor series expansion of $f(z)$.

\subsection{Chebyshev polynomials}

A second possible approach aimed at extending the convergence radius of the cosmographic series and, at the same time, at overcoming the degree of subjectivity in truncating the expansion that may still be present in the Pad\'e method, makes use of Chebyshev polynomials. The latter, in fact, are able to highly reduce the uncertainties on higher-order cosmographic coefficients and, thus, provide an accurate description of the late-time evolution of the universe\cite{chebyshev}. 
Specifically, the Chebyshev polynomials of the first kind are given as
\begin{equation}
T_n(z)=\cos(n\theta)\,, \quad  \theta=\arccos(z)\,, \quad n\in\mathbb{N}
\end{equation}
obeying the recurrence relation $T_{n+1}(z)=2zT_n(z)-T_{n-1}(z)$, such that
\begin{equation*}
\int_{-1}^{1}T_n(z)\ T_m(z)\ w(z)\ dz=
\begin{cases}
\pi\ , & n=m=0 \vspace{0.2cm}\\
\dfrac{\pi}{2} \delta_{nm}\ , & \text{otherwise}
\end{cases}
\end{equation*}
Then, analogously to the Pad\'e technique, one can build rational Chebyshev polynomials as \cite{chebyshev}
\begin{equation}
R_{n,m}(z)=\frac{\displaystyle{\sum_{i=0}^n}\ a_i T_i(z)}{1+\displaystyle{\sum_{j=1}^m}\ b_j T_j(z)}\,.
\label{eq:rational Chebyshev}
\end{equation}

\subsection{Rational approximations vs standard cosmography}

\begin{figure}
\centering
\includegraphics[width=3.2in]{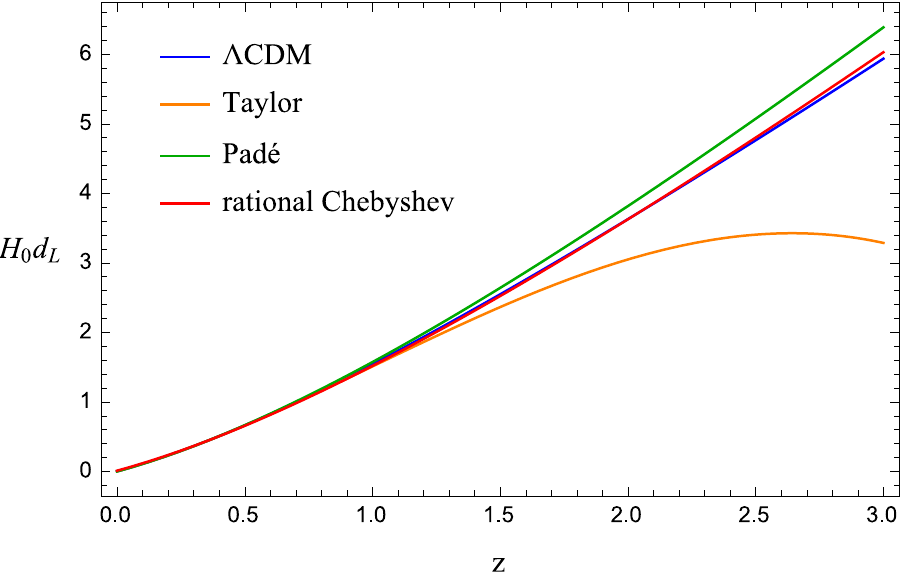}
\caption{4th-order Taylor, (2,2) Pad\'e and (2,1) Chebyshev approximations of the luminosity distance compared with the predictions of the  $\Lambda$CDM model.}
\label{fig:dL_comparison}
\end{figure}

The advantages of rational approximations based on Pad\'e and Chebyshev cosmography with respect to the standard Taylor approach can be verified by testing the effective improvement in terms of stability at high-redshift domains.
For instance, taking into account the reference values obtained for the $\Lambda$CDM model by assuming $\Omega_{m0}=0.3$, namely $(q_0\,,j_0\,,s_0)=(-0.55\,,1\,,-0.35)$, we show  in Figure~\ref{fig:dL_comparison} the significant improvements resulting from the use of the Pad\'e and Chebyshev polynomials as they are able to fairly approximate the $\Lambda$CDM luminosity distance at high $z$, while the accuracy of the Taylor approximation gets worse and worse as soon as $z>1$\cite{chebyshev}.

\section{Cosmographic parametrization of dark energy}

In this section, we provide a first example of application of the cosmographic method. In particular, we shall study the dark energy features in an effective manner  by combining kinematic reconstructions and thermodynamic requirements. 
The constraints coming from the entropy of the universe and the properties of the deceleration parameter may allow, in fact, to investigate the cosmic history without assuming any specific underlying cosmology.

\begin{figure}
\centering
\includegraphics[width=2.3in]{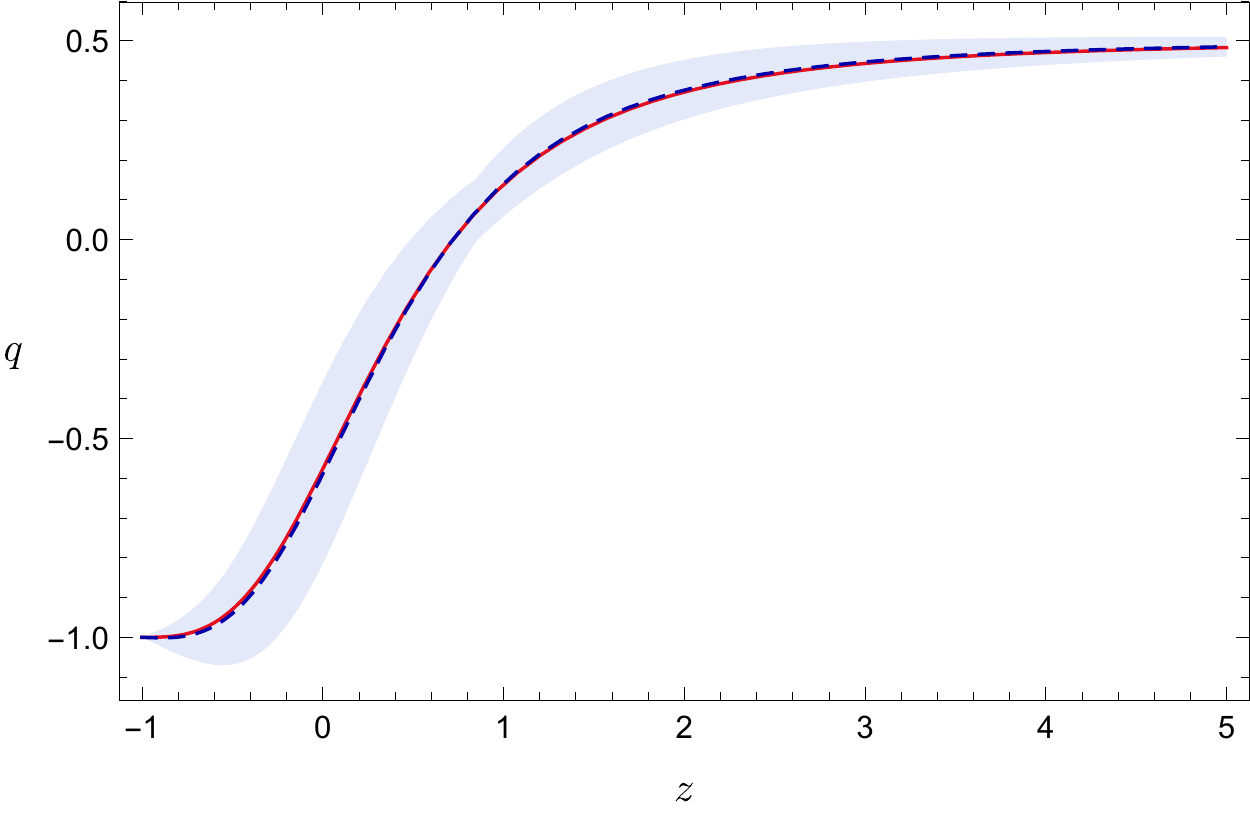}
\includegraphics[width=2.3in]{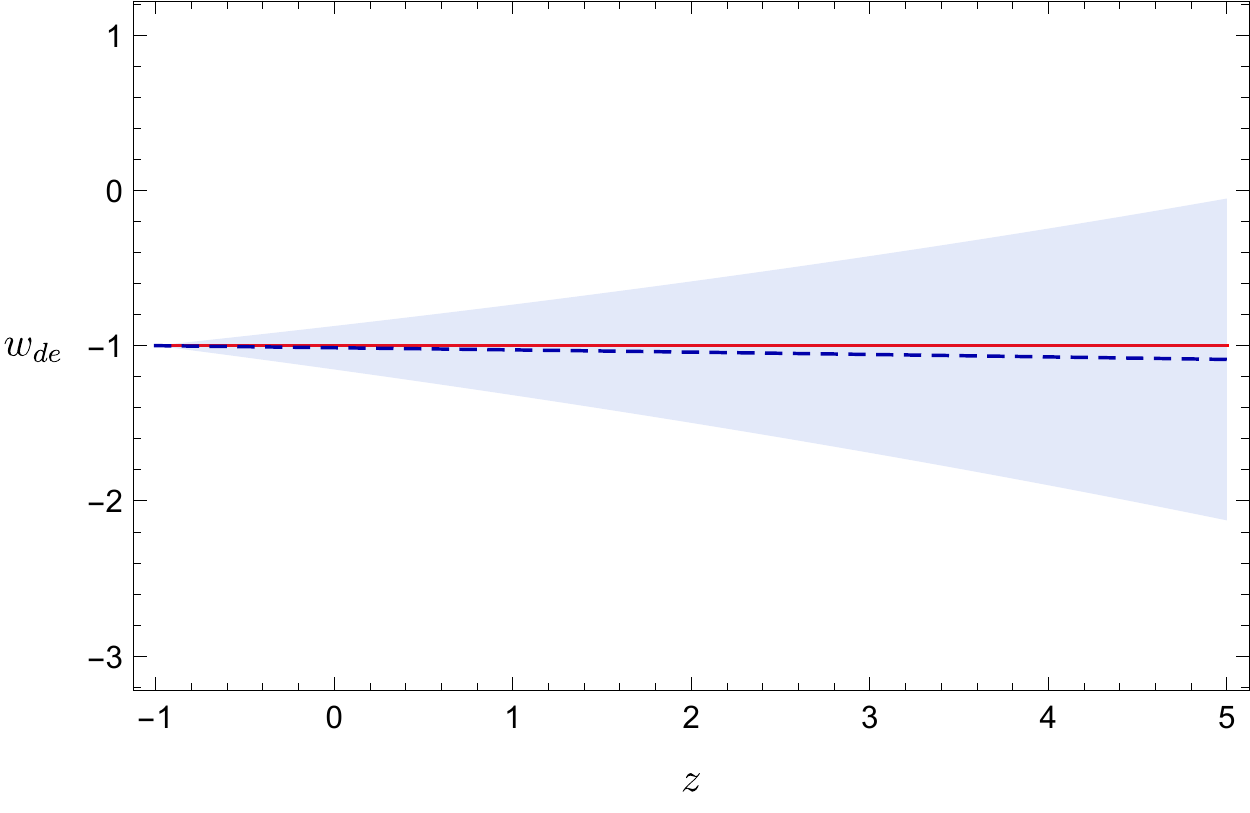}
\caption{$1\sigma$ reconstruction of the deceleration and dark energy equation of state parameters compared to the $\Lambda$CDM predictions (red curves).}
\label{fig:thermo}
\end{figure}

Let us consider the relation between the entropy of the apparent horizon and its area:
\begin{equation}
S_h \propto \mathcal{A}_h=4\pi \tilde{r}_h^2\ , \quad \tilde{r}_h=(H^2+ka^{-2})^{-1/2}\,,
\end{equation}
and the constraints from the second law of thermodynamics: 
\begin{align}
&\mathcal{A}_h'\geq 0\,, \quad \text{at any time} \\
&\mathcal{A}_h''<0\,,\quad  \text{at late times}
\end{align}
Recalling the expression of the deceleration parameter, in the case of vanishing spatial curvature $(k=0)$, we have
\begin{align}
&\dfrac{\mathcal A_h'}{\mathcal A_h}=\dfrac{2}{a}(1+q) \geq 0 \Longrightarrow q\geq-1 \,, \ \forall z \label{eq: first constraint} \\
& \dfrac{\mathcal A_h''}{\mathcal A_h}=\dfrac{2}{a}\left[q'+\dfrac{2q(1+q)}{a}\right] \underset{a\gg 1}{\sim} \frac{2q'}{a} <0 \Longrightarrow \dfrac{dq}{dz}>0 \,, \ z\rightarrow -1 \label{eq: second constraint} 
\end{align}
Moreover, a further constraint could be obtained from the observational predictions of structure formation:
\begin{equation}
q\rightarrow\frac{1}{2}\,, \ z\gg 1\,. 
\label{eq: third constraint} 
\end{equation}
Now, we consider the (0,1) Pad\'e parametrization of the deceleration parameter relative to the dark energy term, 
\begin{equation}
q_{de}(z)=\dfrac{q_{de,0}}{1+q_1 z}\,,
\end{equation}
guaranteeing the sub-dominant behaviour of dark energy with respect to matter: $q_{de}\rightarrow0$ as $z\rightarrow\infty$.
Hence, imposing the constraints \eqref{eq: first constraint}, \eqref{eq: second constraint} and \eqref{eq: third constraint} provides us with a model-independent parametrization of dark energy through the following form of the total deceleration parameter\cite{Capozziello22}:
\begin{equation}
q(z)=\dfrac{2q_{de,0}(1-\Omega_{m0})(1+z+q_{de,0}z)+\Omega_{m0}(1+z)^3}{2\left[(1-\Omega_{m0})(1+z+q_{de,0}z)^2+\Omega_{m0}(1+z)^3\right]}\,.
\end{equation}
The latter leads to 
\begin{equation}
H(z)=H_0\sqrt{\Omega_{m0}(1+z)^3+(1-\Omega_{m0})(1+z+q_{de,0}z)^2}\,,
\end{equation}
which represents a one-parameter extension of the $\Lambda$CDM paradigm, recovered in the limit $q_{de,0}\rightarrow-1$.
Our model can be recast into dynamical dark energy:
\begin{equation}
\left(\dfrac{H}{H_0}\right)^2=\Omega_{m0}(1+z)^3+(1-\Omega_{m0})\exp{\left\{3\int_0^z \dfrac{1+w_{de}(z')}{1+z'}\, dz' \right\}}\,,
\end{equation}
where  the dark energy equation of state parameter is given by
\begin{equation}
w_{de}(z)=\dfrac{1}{3}\left[\dfrac{2q_{de,0}}{1+z(1+q_{de,0})}-1\right].
\end{equation}
The cosmological behaviour of the new dark energy scenario is shown in Figure~\ref{fig:thermo}, as a result of a joint analysis of recent observations\cite{Capozziello22}.

\section{Cosmographic reconstruction of modified gravity}

The cosmographic setups presented above may be employed to reconstruct the gravitational Lagrangian  under different modified gravity frameworks starting from first principles. This  allows us to investigate the origin of dark energy free of possible biases induced by specific cosmological models. In the following, we use units of  $c=8\pi G=1$.

\subsection{The $f(T)$ gravity case}

As first application of the cosmographic method, we analyze the universe's dynamics when gravity is mediated by torsion. The geometry needed to describe spacetime makes use of vierbein fields, $e_A(x^\mu)$, which allow to define the metric tensor as $
g_{\mu\nu}=\eta_{AB}\ {e^A}_\mu\ {e^B}_\nu $, being $\eta_{AB}=\text{diag}(+1,-1,-1,-1)$ the metric of tangent space. Thus, the Lagrangian density of teleparallel equivalent to general relativity (TEGR) can be written as
\begin{equation}
\mathcal{L}_\text{TEGR}=T={S_\rho}^{\mu\nu}{T^{\rho}}_{\mu\nu}\,,
\label{eq:teleparallel lagrangian}
\end{equation}
where $T$ is the torsion scalar, while the torsion tensor is given by $T_{\mu\nu}^\lambda= \hat{\Gamma}_{\mu\nu}^{\lambda}-\hat{\Gamma}_{\nu\mu}^{\lambda}=e^\lambda_A(\partial_\mu e_\nu^A-\partial_\nu e_\mu^A)$, being $\hat{\Gamma}^\lambda_{\mu\nu}=e^\lambda_A \partial_\mu e^A_\nu$ the zero-curvature Weitzenböck connection. Here, we have introduced the  tensor
\begin{equation}
{S_{\rho}}^{\mu\nu}=\dfrac{1}{2}\left({K^{\mu\nu}}_\rho+\delta^\mu_\rho\ {T^{\alpha\nu}}_\alpha- \delta_\rho^\nu\ {T^{\alpha\mu}}_\alpha \right)\,,
\end{equation}
defined in terms of the contortion tensor, ${K^{\mu\nu}}_\rho=-\frac{1}{2}\left({T^{\mu\nu}}_\rho-{T^{\nu\mu}}_\rho-{T_{\rho}}^{\mu\nu}\right)$.

Lagrangian \eqref{eq:teleparallel lagrangian} can be also generalized to include a generic function of the torsion scalar, such that the gravitational action reads
\begin{equation}
S=\int d^4 x\ e \left[\dfrac{f(T)}{2}+\mathcal{L}_m \right],
\end{equation} 
where $e=\sqrt{-g}=\det(e_\mu^A)$, $\mathcal{L}_m$ is the matter field Lagrangian. From the above action, one can derive the following field equations\cite{Bengochea09}:
\begin{equation}
e_A^\rho{S_\rho}^{\mu\nu}(\partial_\mu T)f''+\left[\dfrac{1}{e}\partial_\mu(e e_A^\rho {S_\rho}^{\mu\nu})-e_A^\lambda {T^\rho}_{\mu\lambda}{S_\rho}^{\nu\mu}\right]f'+\dfrac{1}{4}e_A^\nu f=\dfrac{1}{2}e_A^\rho {{T^{(m)}}_\rho}^\nu\,.
\end{equation}
Assuming the flat FLRW line element, such that  $e_A^\mu=\text{diag}(1,a,a,a)$, we obtain the modified Friedmann equations in the form
\begin{align}
H^2&=\dfrac{1}{3}(\rho_m+\rho_T)\,, \\
2\dot{H}+3H^2&=-\dfrac{1}{3}p_T\,,
\end{align}
where non-relativistic matter is assumed to have vanishing pressure and  density $\rho_m=3H_0^2\Omega_{m0}(1+z)^3$, whereas the torsion contribution is accounted for in
\begin{equation}
\rho_T=Tf'(T)-\dfrac{f(T)}{2}-\dfrac{T}{2}\,, \quad p_T=\dfrac{f-Tf'(T)+2T^2f''(T)}{2[f'(T)+2Tf''(T)]}\,.
\end{equation}
Thus, the combination of the Friedmann equations yields\cite{Capozziello17}
\begin{equation}
\left(\frac{df}{dz}\right)^{-1}\left[H (1+z) \frac{d^2f}{dz^2}+3 f \frac{dH}{dz}\right]=\frac{1}{H}\left(\frac{dH}{dz}\right)^{-1}\left[3 \frac{dH}{dz}+(1+z) \frac{d^2H}{dz^2}\right],
\label{eq:master f(T)}
\end{equation}
where we converted the derivatives with respect to $T$ into the derivatives with respect to $z$ through $d/dT=-12H(z)H'(z)d/dz$, following by the relation $T=-6H^2$.
Combining the latter with the first Friedmann equation provides us with the initial condition $f(z=0)=6{H_0}^2(\Omega_{m0}-2)$. Moreover, we may assume no departures of the effective gravitational constant from Newton's value, leading to the second initial condition $df/dz|_{z=0}=1$.

\begin{figure}
\centering
\includegraphics[width=3.1in]{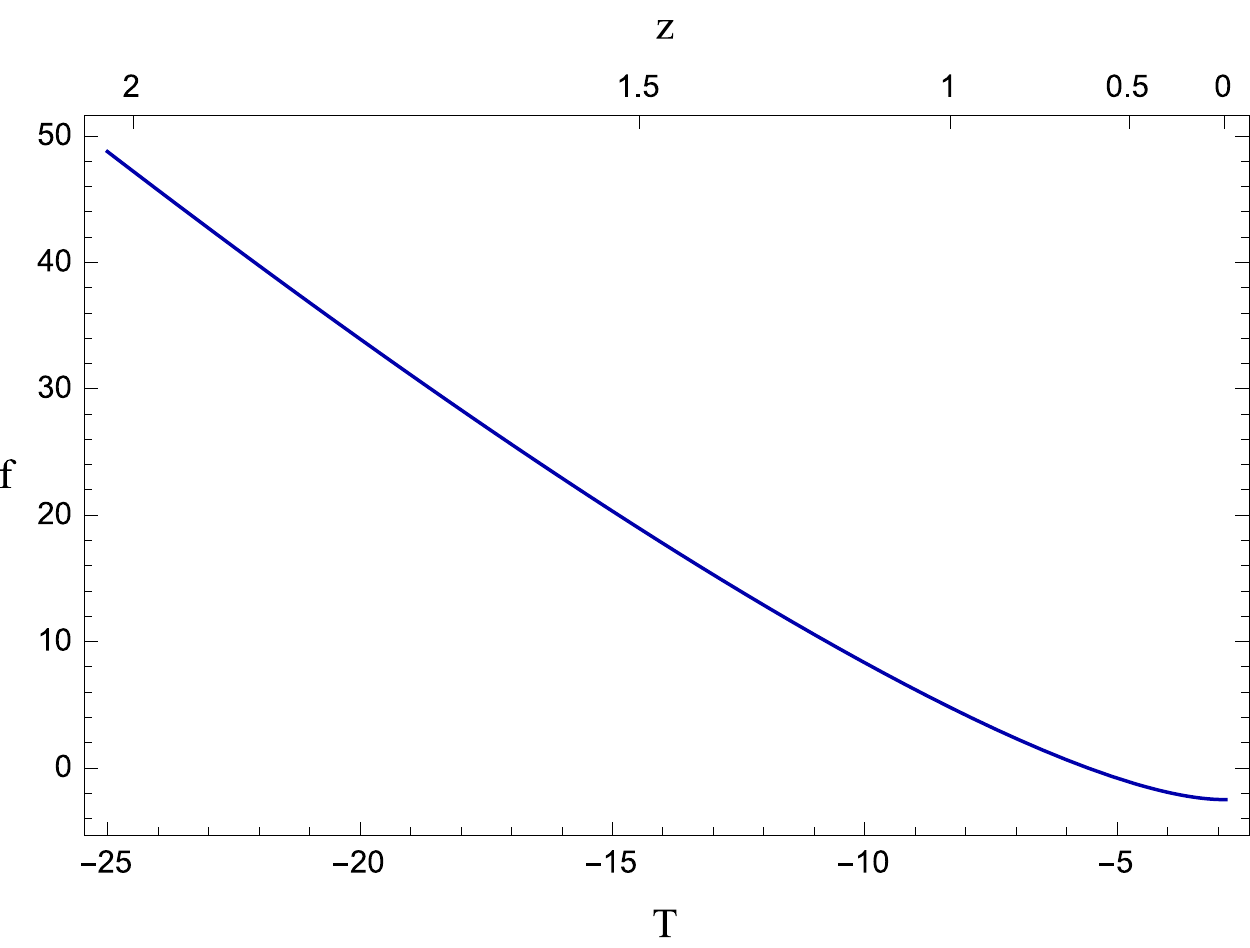}
\caption{Cosmographic reconstruction of $f(T)$ as a function of the redshift based on the 3rd-order Taylor parametrization.}
\label{fig:fT}
\end{figure}

Pursuing a model-independent approach, we can adopt the 3rd-order Taylor approximation given by Eq.~\eqref{eq:H_Taylor} and the best-fit results $(h, q_0, j_0,s_0)=(0.692\,, -0.545\,,0.776\,,-0.192)$, where $h\equiv \frac{H_0}{\text{km/s/Mpc}}$, to find $f(z)$ from numerically integrating Eq.~\eqref{eq:master f(T)}. Thus, after inverting $H(z)$ by means of $T=-6H^2$ to find $z(T)$, the latter can be inserted back into $f(z)$ to finally get the function $f(T)$\cite{Capozziello17}. We show the reconstructed behaviour of $f(T)$ in Figure~\ref{fig:fT}.

\subsection{The $f(R)$ gravity case}

The second application of the cosmographic method we want to discuss concerns $f(R)$ theories, where gravity is induced by non-linear functions of the Ricci curvature, while torsion is vanishing:
\begin{equation}
S=\int d^4x \sqrt{-g}\left[\dfrac{f(R)}{2}+\mathcal{L}_m\right].
\end{equation}
In the metric formalism, the field equations take the form\cite{DeFelice10}
\begin{equation}
R_{\mu\nu}-\dfrac{1}{2}g_{\mu\nu}R=T_{\mu\nu}^{(m)}+\dfrac{1}{f'}\bigg[\dfrac{1}{2}g_{\mu\nu}(f-Rf')+(\nabla_\mu\nabla_\nu-g_{\mu\nu}\square)f'\bigg].
\end{equation}
For a flat FLRW spacetime and non-relativistic matter fields, the Friedmann equations modify as
\begin{align}
&H^2=\dfrac{1}{3}\left[\dfrac{1}{f'}\rho_m+\rho_{curv}\right], \\
&2\dot{H}+3H^2=-p_{curv}\,,
\end{align}
where the effective curvature density and pressure  are give by
\begin{align}
&\rho_{curv}=\dfrac{1}{f'}\left[\dfrac{1}{2}(f-Rf')-3H\dot{R}f''\right].  \\
&p_{curv}=\dfrac{1}{f'}\left[2H\dot{R}f''+\ddot{R}f''+\dot{R}^2f'''-\dfrac{1}{2}(f-Rf')\right].
\end{align}
In this case, the Hubble parameter is related to the Ricci scalar through $R=-6(\dot{H}+2H^2)$, which allows to recast the first Friedmann equation as\cite{Capozziello18}
\begin{align}
&H^2f_z=\Big[-(1+z)H_z^2+H\big(3H_z-(1+z)H_{zz}\big)\Big]\times \Bigg[-6 H_0^2 (1 + z)^3 \Omega_{m0} - f  \nonumber \\
&-\dfrac{Hf_z \left(2 H - (1 + z) H_z\right)}{(1 + z)H_z^2 +H\left(-3H_z+ (1 + z)H_{zz}^2\right)}-\dfrac{(1 + z) H^2}{{\big[(1+z)H_z^2+H\big(-3H_z+(1+z)H_{zz}\big)\big]}^2} \nonumber \\
& \times\Big(f_{zz}\big((1 +z)H_z^2 + H (-3H_z + (1 + z)H_{zz})\big)+f_z\big(2H_z^2-3(1+z)H_zH_{zz} \nonumber \\
&+H(2H_{zz}-(1+z)H_{zzz})\big)\Big) \Bigg],
\label{eq:master f(R)}
\end{align}
where we have converted the derivatives with respect to $R$ into the derivatives with respect to $z$ by means of $\frac{d}{d R}=\frac{1}{6} \Big[(1 + z)H_z^2 +H\left(-3H_z +(1+z)H_{zz}\right)\Big]^{-1}\frac{d}{dz}$. 
Eq.~\eqref{eq:master f(R)} can be solved with the help of the initial conditions obtained from requiring no deviations from the Newton gravitational constant, i.e., $f_0=R_0+6H_0^2(\Omega_{m0}-1)$ and $f_z \big|_{z=0}=R_z\big|_{z=0}$.

\begin{figure}
\centering
\includegraphics[width=3.2in]{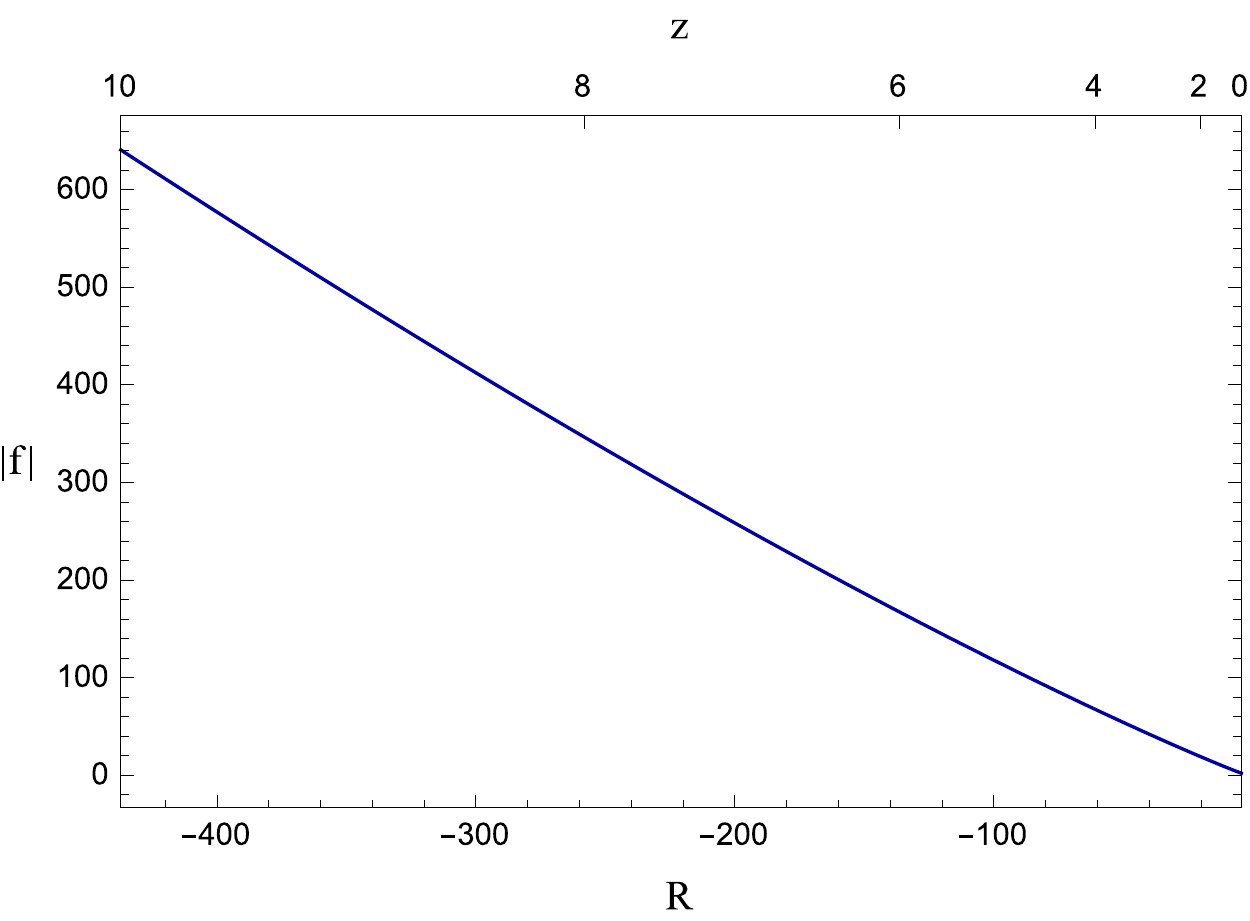}
\caption{Cosmographic reconstruction of $f(R)$ as a function of the redshift based on the (2,1) Pad\'e parametrization.}
\label{fig:fR}
\end{figure}

Motivated by the good properties of Pad\'e polynomials, we consider the (2,1) Pad\'e approximation of the luminosity distance,
\begin{equation}
d(z)\simeq\dfrac{1}{H_0}\left[\dfrac{z (6 (-1 + q_0) + (-5 - 2 j_0 + q_0 (8 + 3 q_0)) z)}{-2 (3 + z + j_0 z) + 2 q_0 (3 + z + 3 q_0 z)}\right] ,
\end{equation}
from which we can derive $H(z)$ to be used in Eq.~\eqref{eq:master f(R)} and find $f(z)$ by means of the best-fit values $(h\,,q_0\,,j_0)=(0.706\,,-0.471\,,0.593)$. Adopting a similar back-scattering procedure as for the $f(T)$ case, one finds that the analytical function that best-approximates the numerical results is\cite{Capozziello18} $f(z)=A+Bz^3 e^{Cz}$, leading to the $f(R)$ function shown in Figure~\ref{fig:fR}. The same procedure can be utilised to reconstruct the shape of $f(R)$ within the Palatini formalism\cite{rocco_Palatini}.

\subsection{The $f(Q)$ gravity case}

A further possibility is to consider gravity as a result of non-metricity, while curvature and torsion are vanishing. In fact, the most general form of affine connections is\cite{Hehl94} $
\Gamma^{\lambda}_{\phantom{\alpha}\mu\nu} =
\left\lbrace {}^{\lambda}_{\phantom{\alpha}\mu\nu} \right\rbrace +
K^{\lambda}_{\phantom{\alpha}\mu\nu}+
 L^{\lambda}_{\phantom{\alpha}\mu\nu} 
$ ,
where $\left\lbrace {}^{\lambda}_{\phantom{\alpha}\mu\nu} \right\rbrace \equiv \frac{1}{2}\,g^{\lambda \beta} \left( \partial_{\mu} g_{\beta\nu} + \partial_{\nu} g_{\beta\mu} - \partial_{\beta} g_{\mu\nu} \right)$ are the Levi-Civita connections. Here, $K^{\lambda}{}_{\mu\nu}$ is the contortion tensor, $L^{\lambda}{}_{\mu\nu} \equiv \frac{1}{2}\, g^{\lambda \beta} \left( -Q_{\mu \beta\nu}-Q_{\nu \beta\mu}+Q_{\beta \mu \nu} \right)$ is the disformation tensor, and $Q_{\rho \mu \nu} \equiv \nabla_{\rho} g_{\mu\nu} = \partial_\rho g_{\mu\nu} - \Gamma^\beta{}_{\rho \mu} g_{\beta \nu} -  \Gamma^\beta{}_{\rho \nu} g_{\mu \beta}$ is the non-metricity tensor\cite{Jarv18}. We can thus consider a gravitational action of the form
\begin{equation}
S=\int  d^4x\, \sqrt{-g}\left[\dfrac{1}{2}f(Q)+\mathcal{L}_m\right],
\end{equation} 
where $f(Q)$ is a generic function of the non-metricity scalar:
\begin{equation}
Q=-\dfrac{1}{4}Q_{\alpha\beta\mu}Q^{\alpha\beta\mu}+\dfrac{1}{2}Q_{\alpha\beta\mu}Q^{\beta\mu\alpha}+\dfrac{1}{4}Q_{\alpha}Q^{\alpha}-\dfrac{1}{2}Q_{\alpha}\tilde{Q}^\alpha\,.
\end{equation}
Hence, the field equations are given as
\begin{align}
&\dfrac{2}{\sqrt{-g}}\nabla_\alpha\bigg\{\sqrt{-g}\, g_{\beta\nu}\, f_Q\Big[-\dfrac{1}{2}L^{\alpha\mu\beta}- \dfrac{1}{8}\left(g^{\alpha\mu}Q^\beta+g^{\alpha\beta}Q^\mu\right)+\dfrac{1}{4}g^{\mu\beta}(Q^\alpha-\tilde{Q}^\alpha)\Big]\bigg\} \nonumber \\
&+f_Q\Big[-\dfrac{1}{2}L^{\mu\alpha\beta}-\dfrac{1}{8}\left(g^{\mu\alpha}Q^\beta+g^{\mu\beta}Q^\alpha\right)+\dfrac{1}{4}g^{\alpha\beta}(Q^\mu-\tilde{Q}^\mu)\Big]Q_{\nu\alpha\beta}+\dfrac{1}{2}{\delta^\mu}_\nu f={T^\mu}_\nu,
\end{align}
where $f_Q\equiv \frac{\partial f}{\partial Q}$. For the line element $ds^2=-dt^2+a(t)^2\delta_{ij}dx^idx^j$, one finds the following modified Friedmann equations\cite{Jimenez20}:
\begin{align}
6H^2f_Q-\dfrac{1}{2}f&=\rho\,, \\
\left(12H^2f_{QQ}+f_Q\right)\dot{H}&=-\dfrac{1}{2}(\rho+p)\,.
\end{align}
Neglecting the pressure of non-relativistic matter, we have $\rho=3H_0^2 \Omega_{m0}(1+z)^{3}$. In the coincident gauge, where general relativity is recovered in the case $f(Q)=Q$, the relation $Q=6H^2$ holds true, implying $f_Q=\frac{f'(z)}{12\,H(z)\,H'(z)}$. Therefore, the first Friedmann equation can be written as\cite{Capozziello22}
\begin{equation}
\dfrac{H'(z)}{H(z)}f'(z)-f(z)=6H_0^2\Omega_{m0}(1+z)^3\,.
\label{eq:master_f(Q)}
\end{equation}
If one requires no deviations from Newton's gravitational constant, then $f_Q|_{z=0}=1$, leading to the initial condition $f_0=6H_0^2(2-\Omega_{m0})$. 

\begin{figure}
\centering
\includegraphics[width=3.2in]{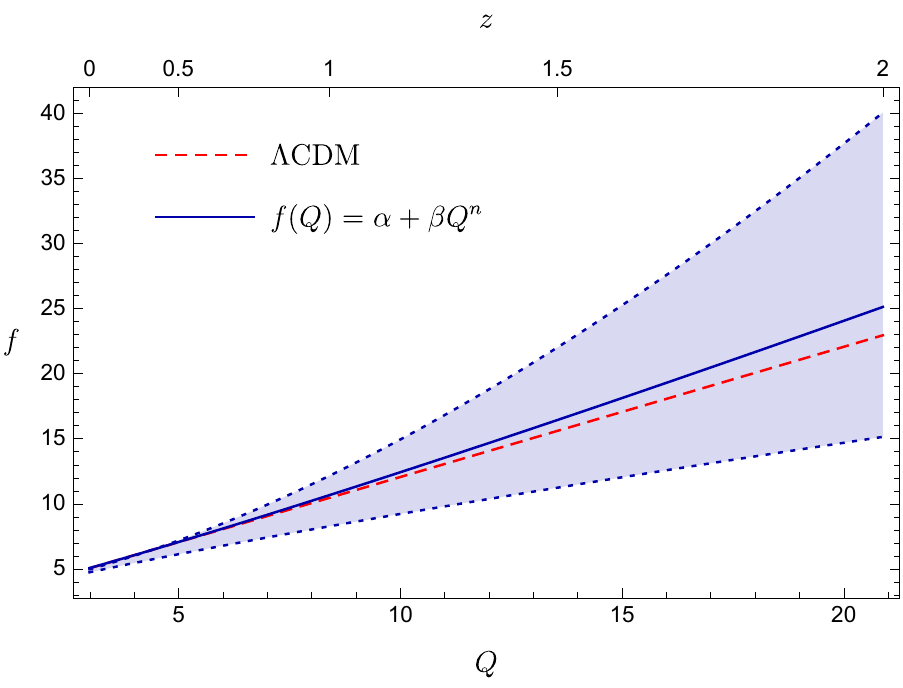}
\caption{Cosmographic reconstruction of $f(Q)$ compared to the predictions of the concorance $\Lambda$CDM cosmology.}
\label{fig:fQ}
\end{figure}

Following the analogous strategy as above, we can use the (2,1) Pad\'e parametrization for $H(z)$ along with the observational constraints $h=0.693\pm 0.002$, $q_0=-0.73\pm 0.13$ and
 $j_0=2.84^{+1.00}_{-1.23}$ to find $f(z)$. Then, we perform a numerical inversion by means of $Q=6H^2$ to finally get $f(Q)$. In so doing, we find that the best analytical matching is provided by the function $f(Q)=\alpha+\beta Q^{n}$, with $(\alpha,\, \beta, \, n)=(2.492,\, 0.757, \, 1.118)$, which  show small departures from the standard $\Lambda$CDM model\cite{Capozziello22}.
However, when taking into account the $1\sigma$ uncertainties on the free parameters, $\alpha \in[2.058,\, 3.162]$, $\beta \in [0.332,\, 1.076]$ and $n \in [0.821,\, 1.550]$, we find no statistically significant deviations with respect to the concordance scenario with $\Omega_{m0}=0.3$ and $h=0.70$ (see Figure~\ref{fig:fQ}).

\section{Final remarks}

The standard description of the universe based on the $\Lambda$CDM cosmology is challenged by long-standing theoretical drawbacks associated to physical interpretation of the cosmological constant and by recent tensions among observations. 
For this reason, the development of model-independent techniques able to feature the late-time acceleration of the universe becomes crucial in order to discriminate among the plethora of cosmological models proposed over the last years as possible solutions to the dark energy problem.
 
In this brief report, we provided an outlook on the current status of the cosmographic method. After reviewing the basic principles of cosmography, we focused on the issues limiting the standard approach in the era of precision cosmology. Thus, we discussed the role of rational parametrizations in view of healing the convergence problem related to Taylor series expansions of cosmological distances when dealing with high-redshift data. 
Specifically, reconstruction techniques relying on Pad\'e and Chebyshev polynomials offer clear advantages in terms of stability and accuracy of cosmographic series including high-order coefficients. 
Rational cosmography can be used to address dark energy from a model-independent perspective. In particular, the combination of thermodynamic principles and Pad\'e modeling of the deceleration parameter provides us with a cosmological scenario that reproduces the $\Lambda$CDM behaviour only under particular limits. This allows us to explore discrepancies with respect to the standard paradigm avoiding biases induced by setting a specific cosmic expansion.

On the other hand, we explored applications of the cosmographic approach to modified gravity scenarios. Under the hypothesis that the current cosmic speed-up may be due to extensions of general relativity, we considered the cosmological consequences of non-linear functions of the Ricci curvature, or gravitational interaction mediated by torsional effects and non-metricity. Thus, we described a back-scattering procedure to reconstruct the $f(R)$, $f(T)$ and $f(Q)$ actions by exploiting the most recent constraints on the cosmographic coefficients. Our results show that the cosmographic method is effective in suggesting the models that better fit to kinematics, although no substantial evidence for deviations from the standard $\Lambda$CDM model is highlighted. This is basically due to the elevated uncertainties affecting the cosmographic coefficients of high order that, currently, do not permit to draw any final conclusions. Therefore, these methods will need to be improved in order to be completely predictive with more stringent observational bounds. This would enable cosmography to shed light on the dark energy nature in view of new and more precise data measurements at high redshifts from future cosmology surveys.

\section{Acknowledgements}
The authors are thankful to Istituto Nazionale di Fisica Nucleare (INFN), Sezione di Napoli, {\it iniziative specifiche} QGSKY and MOONLIGHT2 for financial support.


%
\end{document}